\documentclass[10pt,draftclsnofoot,onecolumn]{IEEEtran}

\usepackage{config}
\usepackage{subfigure}

\author{\IEEEauthorblockN{Abdoulaye Tall\IEEEauthorrefmark{1}, Zwi Altman\IEEEauthorrefmark{1} and Eitan Altman\IEEEauthorrefmark{2}} \\ \IEEEauthorblockA{\IEEEauthorrefmark{1}Orange Labs 38/40 rue du General Leclerc, 92794 Issy-les-Moulineaux \\Email: \{abdoulaye.tall,zwi.altman\}@orange.com}\\ \IEEEauthorblockA{\IEEEauthorrefmark{2}INRIA Sophia Antipolis, 06902 Sophia Antipolis, France, Email:eitan.altman@sophia.inria.fr}
}

\title{Self-optimizing load balancing with backhaul-constrained radio access networks}

\begin{document}
\maketitle

\begin{abstract}

\ac{SON} technology aims at autonomously deploying, optimizing and
repairing the \ac{RAN}. \ac{SON} algorithms typically use \acp{KPI}
from the \ac{RAN}. It is shown that in certain cases, it is
essential to take into account the impact of the backhaul state in
the design of the \ac{SON} algorithm. We revisit the \ac{BS} load
definition taking into account the backhaul state. We provide an
analytical formula for the load along with a simple estimator for both
elastic and \ac{GBR} traffic. We incorporate the proposed load estimator in
a self-optimized \ac{LB} algorithm. Simulation results for a backhaul constrained heterogeneous
network illustrate how the correct load definition can guarantee a
proper operation of the SON algorithm.


\end{abstract}

\section{Introduction} \label{sec:introduction}

The \ac{SON} concept has been introduced by 3GPP
\cite{3gpp_evolveduniversalterrestrial_2008a} as a means to manage
complexity, to reduce cost of operation, and to enhance performance
and profitability of mobile networks. Self organizing networks aim
at autonomously configuring newly deployed network nodes
(self-configuration), at tuning parameters to improve \ac{KPIs}
(self-optimization) and at diagnosing and repairing faulty network
nodes (self-healing). Research on \ac{SON} has mainly focused on the
\ac{RAN} with the assumption of infinite backhaul capacity. However,
finite backhaul capacity may impact the \ac{RAN} performance in
general and the operation of \ac{SON} functions in particular. This
paper investigates self-optimizing \ac{LB} algorithm in the
case of finite backhaul capacity, and proposes solutions to
guarantee correct operation of the algorithm.

Different \ac{SON} algorithms for \ac{LB} have been proposed
in the literature (e.g. \cite{khan_centralizedLB_2013}, \cite{combes_selforganizationwireless_2012}, \cite{nasri_handoveradaptation_2013}, \cite{rodriguez_loadbalancing_2011}).
The \ac{SON} function monitors \acp{KPI} in the \acp{BS} and adjusts
its parameters in order to steer those \acp{KPI} to desired values.
In previous works, the \acp{KPI} are limited to the \ac{RAN}, thus
excluding finite backhaul capacity.

Network operators carefully dimension the backhaul to avoid capacity
bottlenecks and to ensure end-to-end performance, while
avoiding over dimensioning due to both equipment and deployment
costs. In existing networks, and particularly for low power nodes
such as small cells but not only, performance issues due to
finite backhaul capacity can be encountered in \ac{ADSL}
\cite{small_cell_forum_BH}, or in wireless backhaul. In 5G networks,
the high rate requirements for bandwidth intensive services
\cite{osseiran2014scenarios} may cause saturation
even in optical backhaul, unless very costly investment in the
transport network are made. 
%The topic of SON for backhaul as a means
%for managing limited backhaul capacity has recently received
%important interest, mainly by network operators and infrastructure
%vendors, see for example
%\cite{NokiaBackhaulSON},\cite{EricssonBackhaulSON}.

The impact of the backhaul has been considered in non-3GPP networks
e.g. in \ac{LB} problems in \acp{WLAN}
\cite{bejerano_fairness_wlan_2007} where the authors propose load
balancing algorithms but for a static scenario in which the number
of users and their channel conditions are considered fixed. The
backhaul limitations have been studied for LTE but only when it is
used to exchange information between BSs e.g. in Coordinated
Multi-Point (CoMP) \cite{lee_comp_2012}. The backhaul impact has also been studied in
\ac{RRM} mechanisms such as scheduling in
\cite{ghimire_impactbackhaul_2014} but with a static user
association mechanism.

%This paper aims to demonstrate the impact of backhaul limitations on
%a 3GPP SON use case taking into account the dynamics of the system,
%and propose a solution to take that impact into consideration. We 
%consider dynamic traffic and the goal is to
%balance the average loads of \acp{BS} in the network. We show via
%numerical simulations how a \ac{LB} \ac{SON} can fail when
%neglecting the limited backhaul. 
This paper analyses the impact of limited backhaul capacity on 3GPP 
\ac{LB} \ac{SON} taking into account the system dynamics. 
Numerical simulations show how a \ac{LB} algorithm can fail when 
neglecting the limited backhaul.
The contributions of the paper are the following:
\begin{itemize}
\item A global load definition for a \ac{BS} taking into
account the traffic demand and the capacity of both the backhaul and \ac{RAN}.
\item A simple and measurable estimator for the global load.
\item Simulation results that show the limits of state-of-the-art load
balancing algorithms in backhaul-constrained settings and the way
these limits can be overcome using the global load indicator.
\end{itemize}

The rest of the paper is organized as follows. A classical
definition of the load and a \ac{LB} algorithm
are recalled in Section \ref{sec:legacyLB}. The corrected load
balancing algorithm is presented in Section \ref{sec:BHcontrainedLB}
along with the modified load definition which takes into account the
backhaul state. Section \ref{sec:numerical_illustration} describes
the numerical results which highlight the importance of using the
correct load estimator to avoid significant performance
deterioration. Section \ref{sec:conclusion} concludes the paper.

\section{\ac{LB} For Infinite-Backhaul} \label{sec:legacyLB}

Consider the downlink of a mobile network such as the \ac{LTE}. We
suppose that the backhaul has an infinite capacity or at least
greater than the capacity of the \ac{BS}. Two equivalent definitions
can be used for the load of the \ac{BS}: The first is  the
occupation rate of its resources. The second is the ratio between
the traffic demand and the cell capacity which is valid for both
elastic or \ac{GBR} traffic.
%The second is the ratio between the traffic demand, be it
%elastic or \ac{GBR}, and the cell capacity.

In the case of a \ac{BS} serving elastic traffic, users arrive
randomly according to a Poisson process of intensity $\lambda(r)$ (in users/s/m$^2$) at position $r$,
download a file of random size $\sigma$ with mean $\mathbb{E}(\sigma)$ (in Mbits) and
leave the network when their download is complete. The load is
written as \cite{bonald_wirelessdownlinkdata_2003a}

    \begin{equation} \label{eq:load_bonald}
    \rho_e = \min \left(1, \int_A \frac{\lambda(r) \mathbb{E}(\sigma)}{R(r)} dr \right),
    \end{equation}
where $A$ is the area of the considered cell, 
and $R(r)$ (in Mbps) is the peak data rate (i.e. when the user is 
alone in the cell) at position $r$ averaged over fading.

If the \ac{BS} also serves \ac{GBR} traffic with fixed amount of
resources allocated to the users, the priority is
given to the \ac{GBR} users. 
%We assume that \ac{GBR} users' arrivals
%and departures occur at the same time scale as the load evaluations
%thus the \ac{GBR} traffic can be considered as constant. 
We consider that \ac{GBR} traffic varies slowly with respect 
to elastic traffic (which is bursty) and is considered constant in \eqref{eq:load_local_analytic}.
The load is redefined as

    \begin{equation} \label{eq:load_local_analytic}
    \rho = \min \left(1, \rho_{GBR} + \int_A \frac{\lambda(r) \mathbb{E}(\sigma)}{\bar{R}(r)} dr \right),
    \end{equation}
where $\bar{R}(r)$ is the peak data rate achievable at position $r$
when using only the resources left after scheduling all \ac{GBR}
users, and $\rho_{GBR}$ is the proportion of resources occupied
by the \ac{GBR} traffic at the radio access (in this case $\bar{R}(r) = (1 - \rho_{GBR}) R(r)$). 
%Consider for example a case where the total available
%bandwidth is 10MHz, and with two \ac{GBR} users each requesting
%7.5Mbps of data rate. We suppose that the spectral efficiency for
%both users is 3.75 bit/Hz/s, then 4MHz of bandwidth (or the
%equivalent amount of resource blocks) is required for the \ac{GBR}
%users, where $\bar{R}(r)$ is the peak data rate achievable with the
%remaining 6MHz. If the resources available cannot satisfy all
%\ac{GBR} users then the load is 1. 
It is noted that this definition
of the load does not depend on the scheduling algorithm which
only impacts the user performance.

    In practice, the load is estimated by the proportion of time-frequency resources that
are occupied by the scheduler over a certain time period. We denote
by $K$ the total number of resource blocks available at a \ac{LTE}
\ac{BS}, by $K_t$ the number of resource blocks used at time slot
$t$, and by $T$ the total number of time slots over which the load is
estimated. The load estimator is then given as

    \begin{equation} \label{eq:load_local_estimate}
    \hat{\rho} = \sum_{t=1}^T \frac{K_t}{K \cdot T}.
    \end{equation}
\ac{LB} consists in updating certain \ac{RRM} or system parameters in order to balance the load across \acp{BS} in the network.

We consider here a \ac{LB} algorithm proposed in
\cite{combes_selforganizationwireless_2012} that tunes the pilot
powers of \acp{BS} in order to adjust the coverage of cells
according to their loads. This algorithm has been developed for
distributed and reactive operation, although it can be adapted for a
centralized \ac{SON} operation. The algorithm is presented in the
form of a \ac{SA} update equation as follows:

    \begin{equation} \label{eq:lb_classic}
    P_s[t+1] = P_s[t] + \epsilon (\rho_0[t] - \rho_s[t]).
    \end{equation}
$P_s$ is the pilot power of \ac{BS} $s$, $\rho_s[t]$ - its load
at time $t$, $\rho_0[t]$ - the load of the reference cell at time
$t$, and $\epsilon$ - a constant step size. The reference \ac{BS} can
be chosen to be the most loaded cell in the considered area.

    The authors in \cite{combes_selforganizationwireless_2012} have 
shown that as $\epsilon \to 0$ and $t \to +\infty$, $P_s$ in \eqref{eq:lb_classic} 
converges in probability to a set of pilot powers for which the
loads of all the \acp{BS} are balanced on the average. Their proof
relies on the elastic traffic scenario but the algorithm remains
valid for \ac{GBR} traffic as well. This \ac{LB} algorithm
has been extended in \cite{tall_selforganizingstrategies_2014} to
heterogeneous network scenario where each macro cell is surrounded by a
number of small cells. The pilot powers are replaced with the
\ac{CIO} of the small cells and the reference cell is chosen as the
nearest macro cell. The \ac{CIO} is used together with the \ac{RSRP} to define the attachment rule for the \ac{UE}
\begin{equation}
s^* = \text{argmax}_s \text{CIO}_s h_s^u P_s,
\end{equation}
where $s^*$ is the chosen serving cell, $\text{CIO}_s$ - the \ac{CIO} of cell $s$ and $h_s^u$ - the pathloss from \ac{BS} $s$ to \ac{UE} $u$.

    %The \ac{LB} algorithm \eqref{eq:lb_classic} can use the loads defined in
%\eqref{eq:load_bonald}, \eqref{eq:load_local_analytic} or
%\eqref{eq:load_local_estimate}. %These load definitions do not take
%into consideration the backhaul capacity which may be required for the algorithm to operate properly. %In the following, we propose new definitions for the loads that can be used by the \ac{LB} algorithm in the case of backhaul-constrained settings.

\section{\ac{LB} With Limited-Backhaul} \label{sec:BHcontrainedLB}

    In order to take into account the impact of backhaul capacity
on \ac{LB} algorithms, the \ac{BS} load definition should be
modified to include the backhaul occupancy. Indeed, when the
backhaul is saturated while the \ac{BS} capacity remains
sufficient, \acp{KPI} such as outage probability or \ac{FTT} may
drastically deteriorate. In this case, the buffer of the \ac{BS}
may be empty since the radio link traffic flows faster than the backhaul
traffic feeding the buffer.

    The load \eqref{eq:load_bonald} for elastic traffic is rewritten taking into account the state of the backhaul as follows

    \begin{equation} \label{eq:load_global_bonald}
    \begin{split}
    \rho_{e,g} = \min \left(1, \int_A \frac{\lambda(r) \mathbb{E}(\sigma)}{\min(C_{BH},R(r))} dr \right)
    \end{split}
    \end{equation}
    where $C_{BH}$ is the capacity of the backhaul reserved for the \ac{RAN} traffic. The subscript $g$
stands for \emph{global}, taking into account both \ac{BS} and
backhaul, as opposed to \emph{local}. %It is noted that we assume $0/0= 1$.

The rationale behind Equation \eqref{eq:load_global_bonald} is
that the limited backhaul capacity may limit the peak data rate of a
\ac{UE} when alone in the cell. Hence Equation
\eqref{eq:load_bonald} should be modified by replacing $R(r)$ with
$min(C_{BH},R(r))$. This modification is validated through
simulation results in Section IV where we compare the adjusted load
formula (in dashed lines in Figs. \ref{fig:loadsl} and
\ref{fig:loadsg}) with the actual load observed in the simulated
system (plain lines in the Figures).

    The load definition in \eqref{eq:load_global_bonald} 
    is modified if \ac{GBR} traffic is also
considered as follows

\begin{equation} \label{eq:load_global_analytic}
    \begin{split}
    \rho_g = \min \left(1, \rho_{GBR,g} + \int_A \frac{\lambda(r) \mathbb{E}(\sigma)}{\min(\bar{C}_{BH},\bar{R}(r))} dr \right).
    \end{split}
\end{equation}
$\bar{C}_{BH}$ is the remaining backhaul capacity after backhaul
resources have been allocated to \ac{GBR} traffic. If we denote by $D_\text{GBR}$ the total traffic demand of \ac{GBR} users then $\bar{C}_{BH} = \max(0,C_{BH} - D_\text{GBR})$. $\rho_{GBR,g} = \max(\frac{D_\text{GBR}}{C_{BH}}, \rho_{GBR})$ is the global load of \ac{GBR} traffic.

    \subsection*{Load estimator}
    In practice, a simple load estimator can be derived, based on scheduler
measurements. Consider first only elastic traffic and assume that
all the resources are occupied even if only one user is
present. Then the load can be estimated by the proportion of time
that at least one user is present in the cell:
    \begin{equation} \label{eq:load_global_estimate}
    \rho_e = \sum_{t=1}^T \frac{\textbf{1}_{\{x_e(t) > 0\}}}{T}
    \end{equation}
    where $x_e(t)$ is the number of users at time slot $t$.

    %    If we consider only \ac{GBR} traffic, the load is given by the average
%over time of the maximum of the resource utilization (i.e. the ratio between the used and total available resources) in the \ac{RAN} ($\gamma$) and in the backhaul ($\rho_{BH}$):
%
%    \begin{equation} \label{eq:gbr_load}
%    \rho_{gbr} = \frac{1}{T} \sum_{t=1}^T \max(\gamma(t), \rho_{BH}(t))
%    \end{equation}
%%    where $\gamma(t)$ is the proportion of resources used by the \ac{GBR} traffic
%%at time slot $t$ in the \ac{RAN}, and $\rho_{BH}(t)$ is the occupancy of the
%%backhaul at time slot $t$.
%
%Recalling that $C_{BH}$ is the total backhaul
%capacity and assuming that the unused backhaul capacity denoted by
%$C_{BH,\text{available}}(t)$ is known, then $\rho_{BH}(t)$ can be
%written as
%
%    \begin{equation} \label{eq:rho_bh}
%    \rho_{BH}(t) = \frac{C_{BH} - C_{BH,\text{available}}(t)}{C_{BH}}
%    \end{equation}
%
%    Combining \eqref{eq:load_global_estimate} and \eqref{eq:gbr_load} for a mixed traffic
%scenario with priority given to \ac{GBR} traffic, we get a general
%load estimator $\hat{\rho}_g$ that reads

    If we consider a mixed traffic scenario with priority 
    given to \ac{GBR} traffic, we get a general
load estimator $\hat{\rho}_g$ that reads
    \begin{equation} \label{eq:load_estimator_global}
    \hat{\rho}_g = \frac{1}{T} \sum_{t=1}^T \textbf{1}_{\{x_e(t) > 0\}} + \mathbf{1}_{\{x_e(t) = 0\}} \max(\gamma(t), \rho_{BH}(t))
    \end{equation}
where $\gamma(t)$ is the proportion of resources used by the \ac{GBR} traffic
at time slot $t$ in the \ac{RAN}, and $\rho_{BH}(t)$ is the occupancy of the
backhaul at time slot $t$.

    The \ac{LB} algorithm is then rewritten as follows
    \begin{equation} \label{eq:lb_global}
    P_s[t+1] = P_s[t] + \epsilon (\hat{\rho}_{g,0}[t] - \hat{\rho}_{g,s}[t])
    \end{equation}
where $\hat{\rho}_{g,s}$ and $\hat{\rho}_{g,0}$ are the global loads of cell $s$
and the reference cell $0$ respectively evaluated using Eq.
\eqref{eq:load_estimator_global}. It is noted that the convergence proof of \eqref{eq:lb_global} is the same as in \cite{combes_selforganizationwireless_2012}.

\section{Numerical results} \label{sec:numerical_illustration}
\begin{figure}[!ht]
\centering
\includegraphics[width=4in]{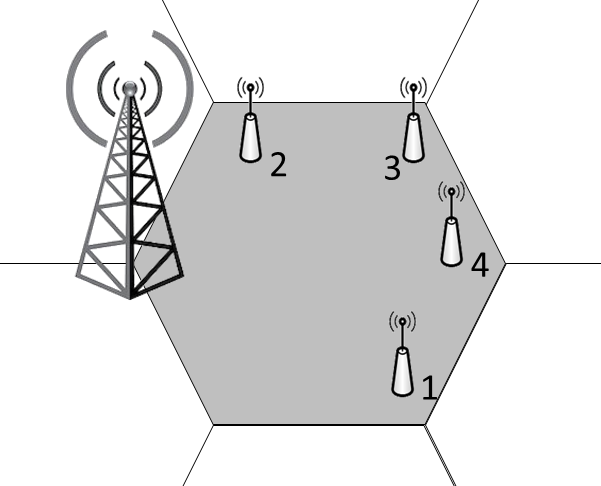}
\caption{Network Layout} \label{fig:net_lay}
\end{figure}

    Consider a \ac{LTE} network comprising a trisector macro \ac{BS} surrounded by 6 interfering
macro \acp{BS}. We select one sector and place in it 4 small cells
(see Fig. \ref{fig:net_lay}). We consider only elastic traffic and
evaluate the performance for the selected macro sector and the small
cells inside its coverage area.

    Two layers of traffic are superposed: the first one has a uniform
arrival rate of $\lambda$ users/s in the entire area (grey area in
Fig. \ref{fig:net_lay}). The second one has a uniform arrival rate
of $\lambda_h$ users/s in the initial area covered by the small
cells (with all \acp{CIO} set to 0dB), namely the small cells are
deployed to serve the users in the hotspot areas.

    To illustrate the impact of a bottleneck at the backhaul,
we assume a low backhaul capacity of 10 Mbps. The propagation models
for the macro \acp{BS} and the small cells (following \cite[Page
61]{3gpp_evolveduniversalterrestrial_2010}) are presented in Table
\ref{tab:params} which also summarizes the simulation parameters.

\begin{table}[!t]
\renewcommand{\arraystretch}{1.3}
\caption{Network and Traffic Parameters} \label{tab:params}
\centering
\begin{tabular}{|c|c|}
\hline
\multicolumn{2}{|c|}{Network parameters} \\
\hline
Number of cells & 1 macro sector, 4 small cells \\
%\hline
%Number of macro sectors & 1 \\
%\hline
%Number of small cells & 4 \\
\hline
Number of interfering macros & 6 $\times$ 3 sectors \\
\hline
Macro Cell layout & hexagonal trisector \\
\hline
Small Cell layout & omni \\
\hline
Intersite distance & 500 m \\
\hline
Bandwidth & 20MHz \\
\hline
\multicolumn{2}{|c|}{Channel characteristics} \\
\hline
Thermal noise & -174 dBm/Hz \\
\hline
Macro Path loss (d in km) & 128 + 36.4 $\log_{10}(d)$ dB \\
\hline
small cell Path loss (d in km) & 140.7 + 36.7 $\log_{10}(d)$ dB \\
\hline
\multicolumn{2}{|c|}{Traffic characteristics} \\
\hline
Traffic spatial distribution & uniform \\
\hline
$(\lambda, \lambda_h)$ & $\text{(8,4) users/s}$ \\
%\hline
%$\lambda$ & $\text{8 users/s}$ \\
%\hline
%$\lambda_h$ & $\text{4 users/s}$ \\
\hline
Service type & FTP \\
\hline
Average file size & 4 Mbits \\
\hline
\end{tabular}
\end{table}

\begin{figure}[ht]
\centering
\subfigure[Local SON]{%
\includegraphics[width=3in]{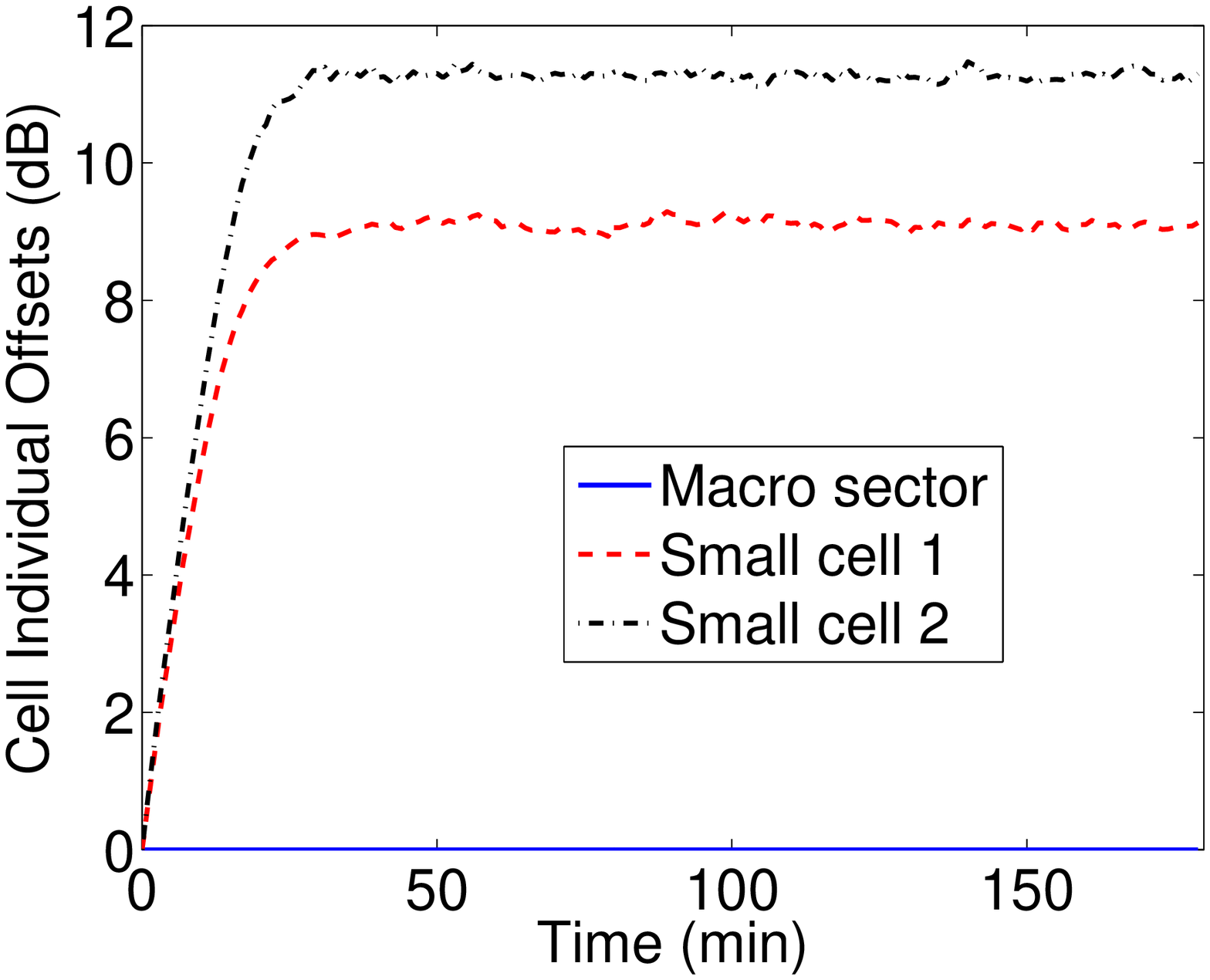}
\label{fig:ciol}}
\quad
\subfigure[Global SON]{%
\includegraphics[width=3in]{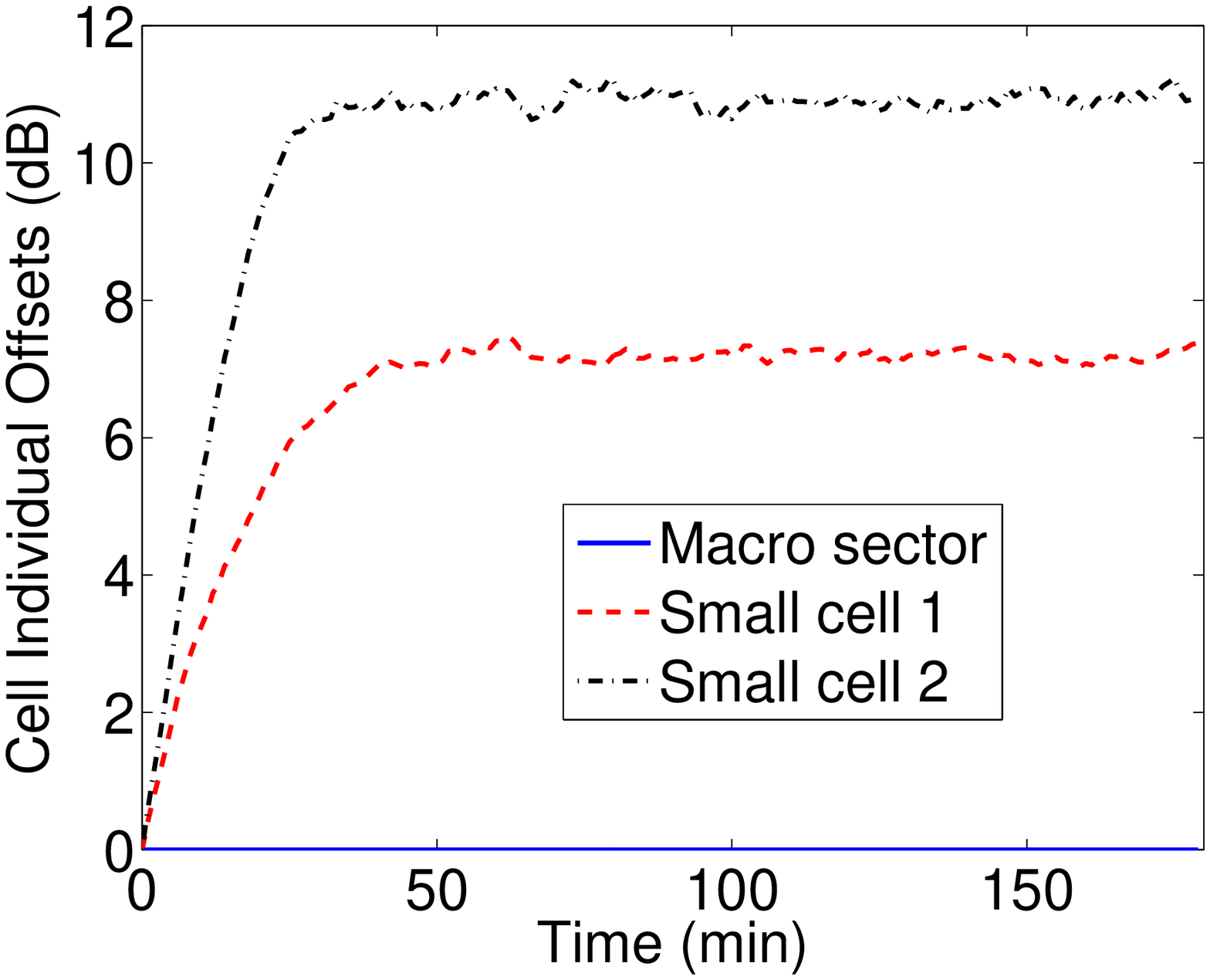}
\label{fig:ciog}}
\caption{Cell Individual Offsets}
\label{fig:cios}
\end{figure}

\begin{figure}[ht]
\centering
\subfigure[Local SON]{%
\includegraphics[width=3in]{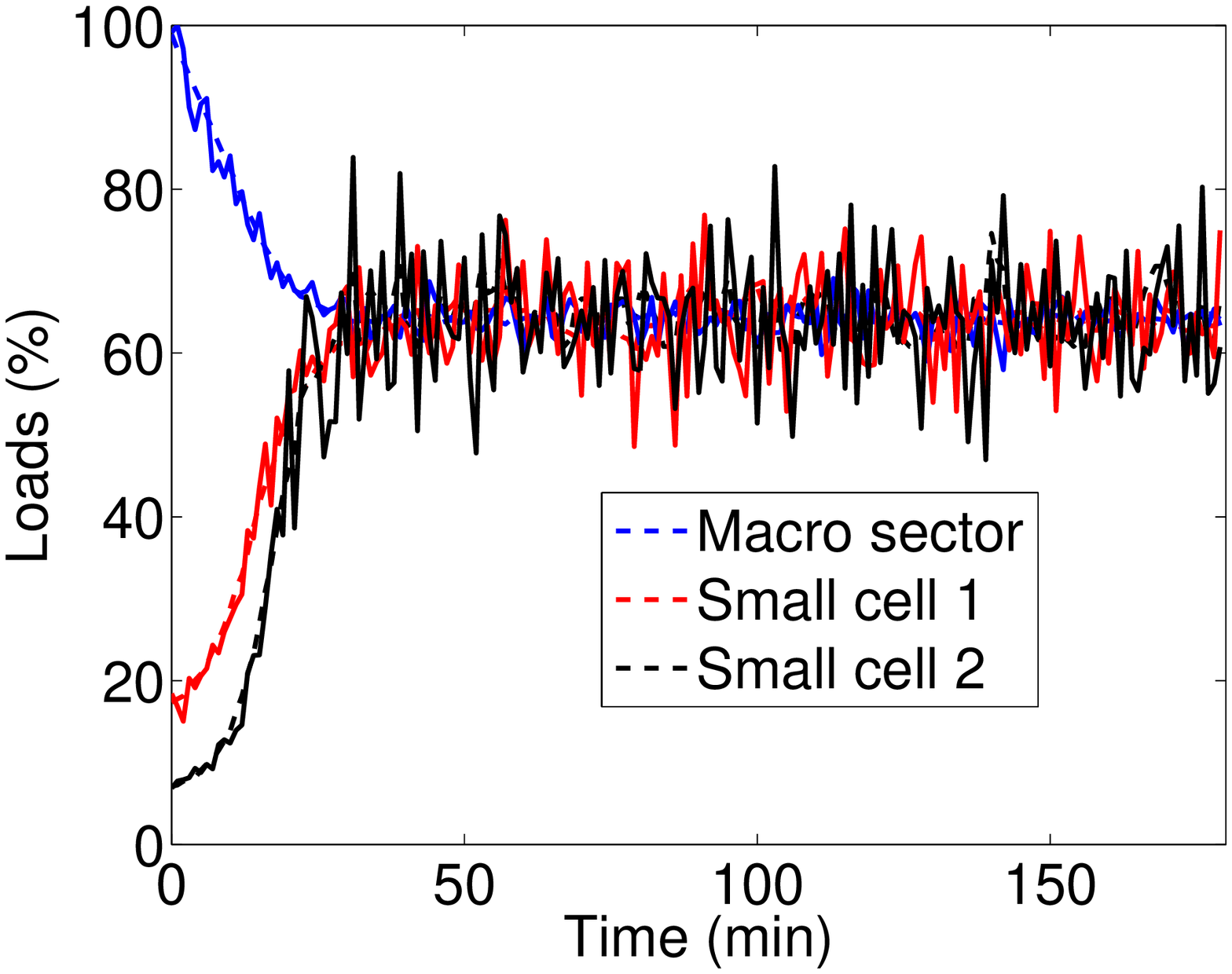}
\label{fig:loadll}}
\quad
\subfigure[Global SON]{%
\includegraphics[width=3in]{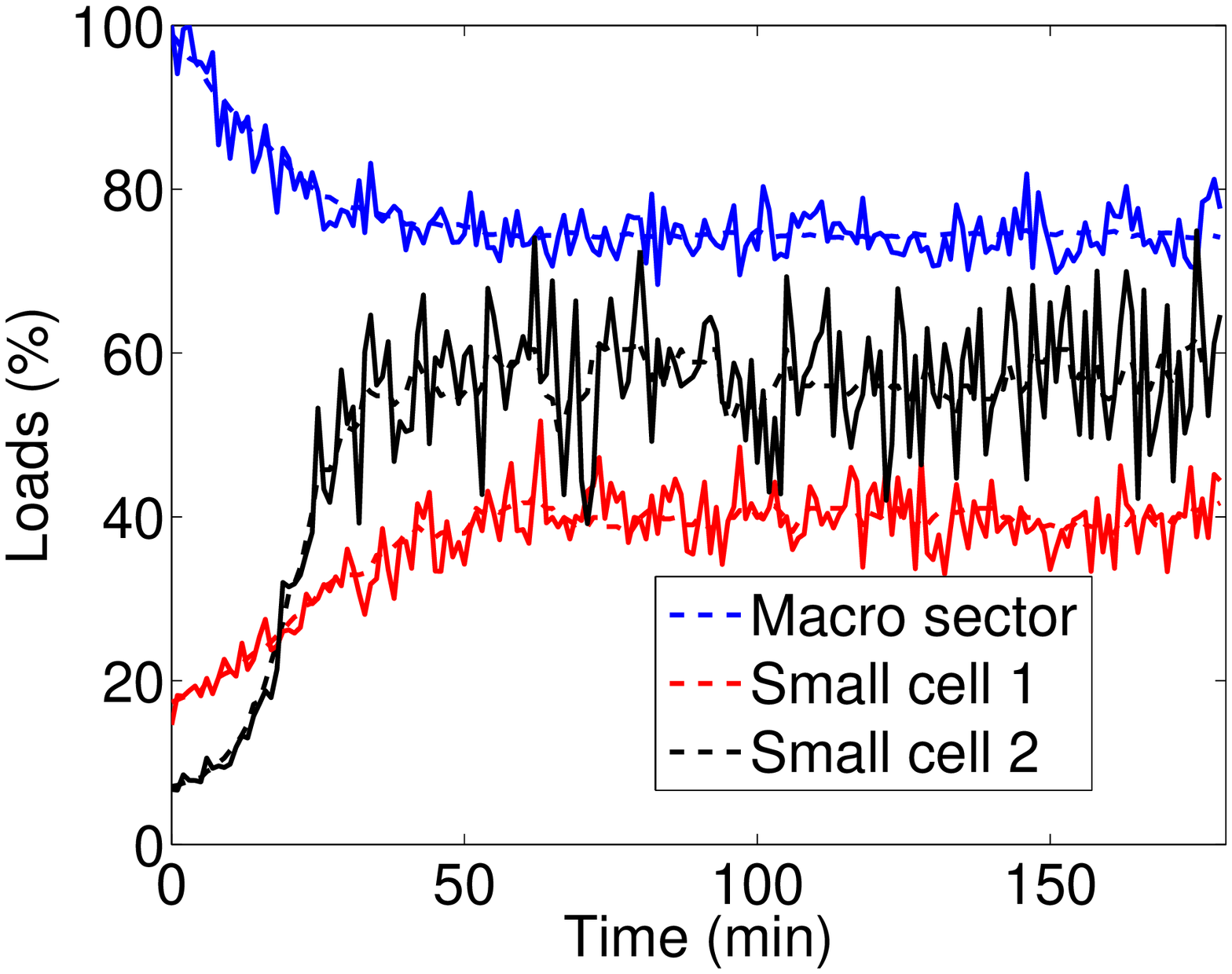}
\label{fig:loadlg}}
\caption{Local loads using Eq. \eqref{eq:load_local_analytic} (dashed lines) and Eq. \eqref{eq:load_local_estimate} (plain lines)}
\label{fig:loadsl}
\end{figure}

\begin{figure}[ht]
\centering
\subfigure[Local SON]{%
\includegraphics[width=3in]{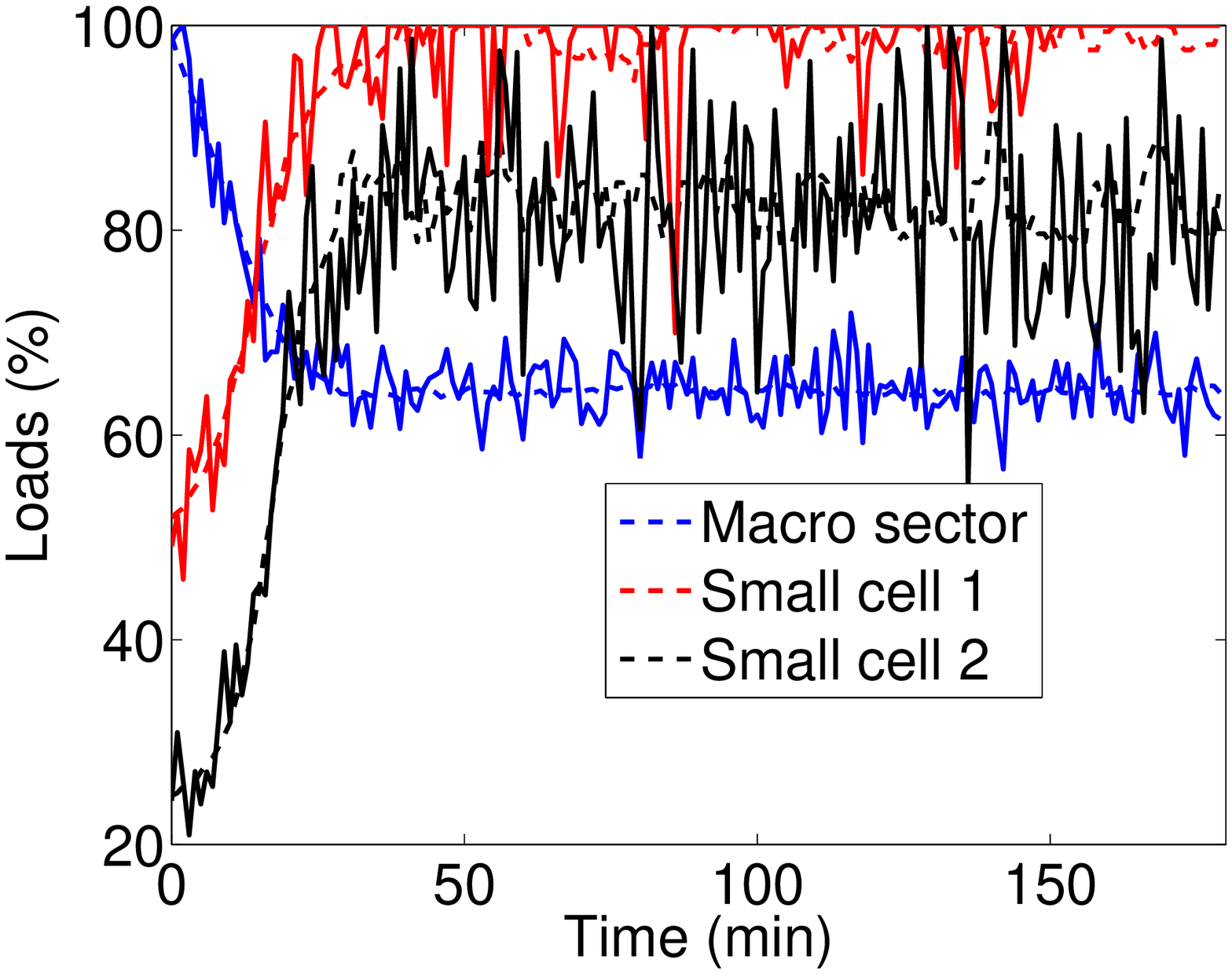}
\label{fig:loadgl}}
\quad
\subfigure[Global SON]{%
\includegraphics[width=3in]{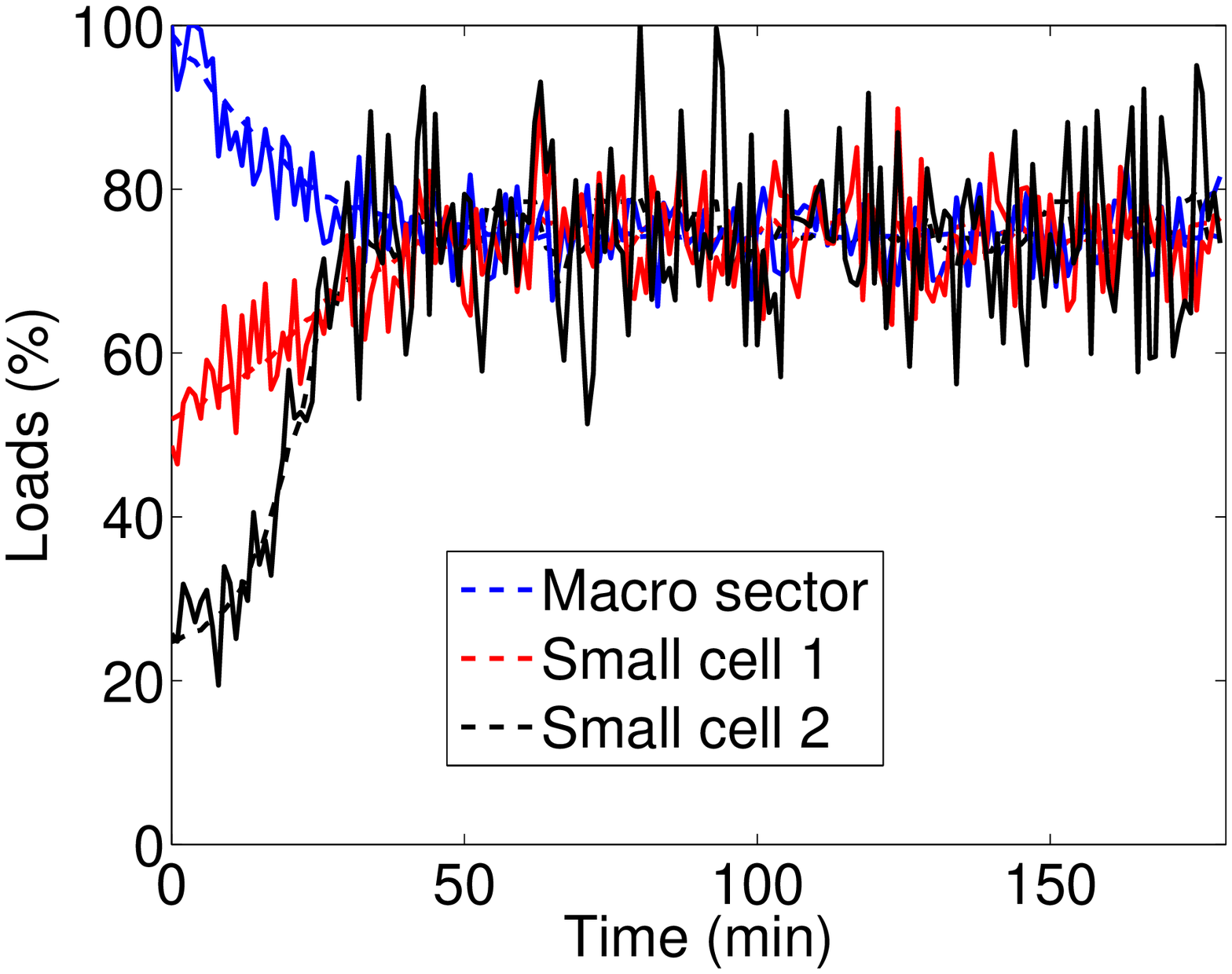}
\label{fig:loadgg}}
\caption{Global loads using Eq. \eqref{eq:load_global_bonald} (dashed lines) and  Eq. \eqref{eq:load_global_estimate} (plain lines)} \label{fig:loadsg}
\end{figure}

\begin{figure}[ht]
\centering
\subfigure[Local SON]{%
\includegraphics[width=3in]{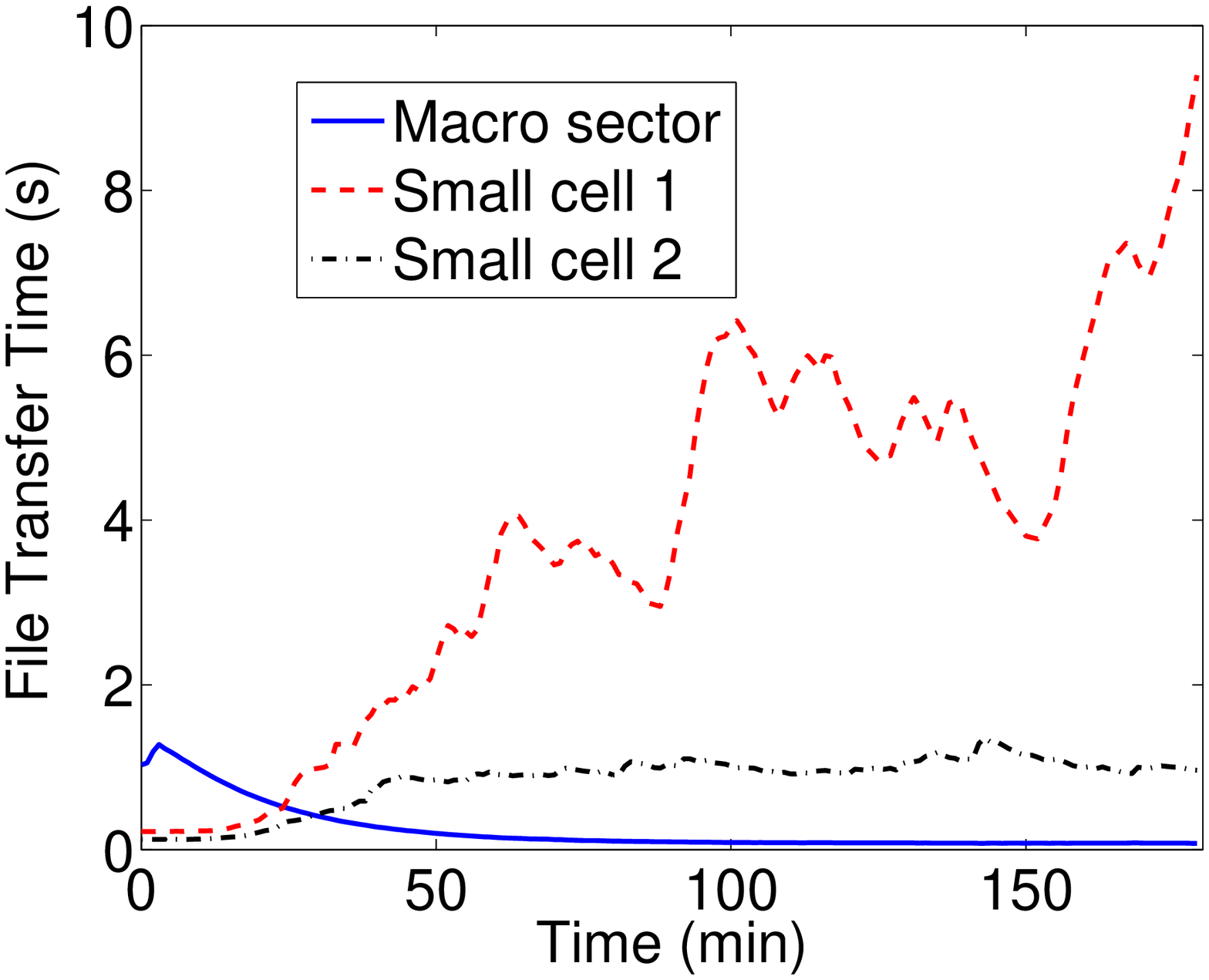}
\label{fig:fttl}}
\quad
\subfigure[Global SON]{%
\includegraphics[width=3in]{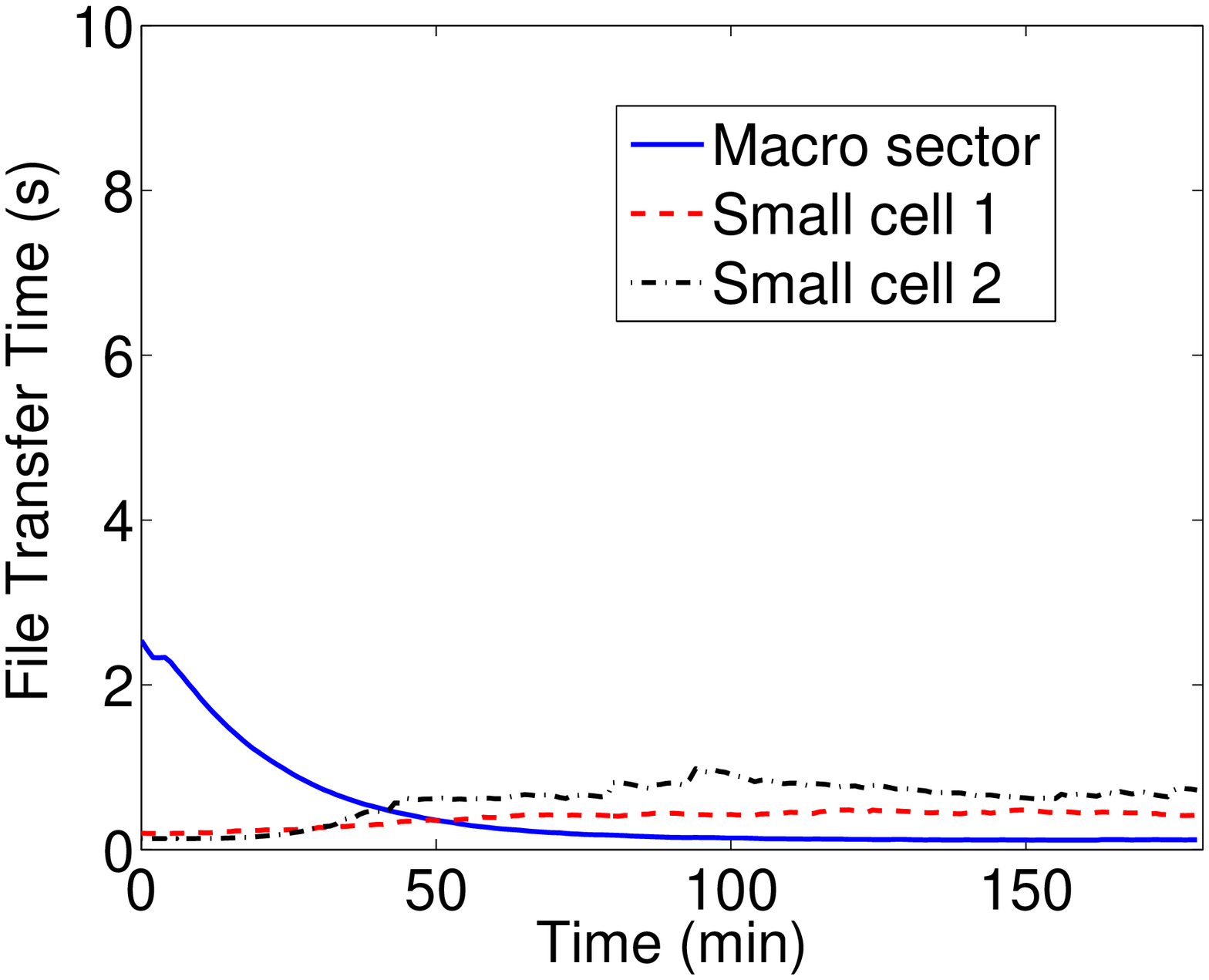}
\label{fig:fttg}}
\caption{File Transfer Time } \label{fig:ftts}
\end{figure}

    We simulate the system during 3 hours and compare
side-by-side the performance obtained using Algorithms
\eqref{eq:lb_classic} and \eqref{eq:lb_global}
denoted respectively as \textit{Local SON} and
\textit{Global SON}. We present the results for the macro sector and for
two of the small cells. We plot the analytical and the
estimated loads (with $T = 60$s) in dashed and plain lines respectively,
using the different definitions (see Figs. \ref{fig:loadsl}
and \ref{fig:loadsg}). We also plot the \acp{CIO} (Fig. \ref{fig:cios})
set by the algorithms over time and the corresponding \acp{FTT} (Fig. \ref{fig:ftts}).

    The local SON balances the \acp{BS}'
scheduler loads (see Fig. \ref{fig:loadll}) while it is unable
to balance the real loads (see Fig. \ref{fig:loadlg}).
In particular, small cell 1 increases excessively its coverage
area (see red curve in Fig. \ref{fig:ciol}) which causes its
\ac{FTT} to explode (red curve in Fig. \ref{fig:fttl}).

    On the other hand, the Global SON balances the real
loads (see Fig. \ref{fig:loadgg}) by limiting the increase in
small cells' \acp{CIO} (see Fig. \ref{fig:ciog}). As a consequence,
the small cells' \ac{FTT} remain low
while the macro cell's \ac{FTT} is decreased
(see Fig. \ref{fig:fttg}).

It is noted that the size of small cell 2 is initially small because of
its proximity to the macro cell. So the increase in its \ac{CIO}
does not increase too much its size, thus its performance
remains good even with local SON as shown in Fig. \ref{fig:fttl}.
%Figs. \ref{fig:loadsl} and
%\ref{fig:loadsg} also show the accuracy of the proposed load
%estimators (plain lines) compared to the analytical loads (dashed
%lines).

\begin{figure}[ht]
\centering
\includegraphics[width=4in]{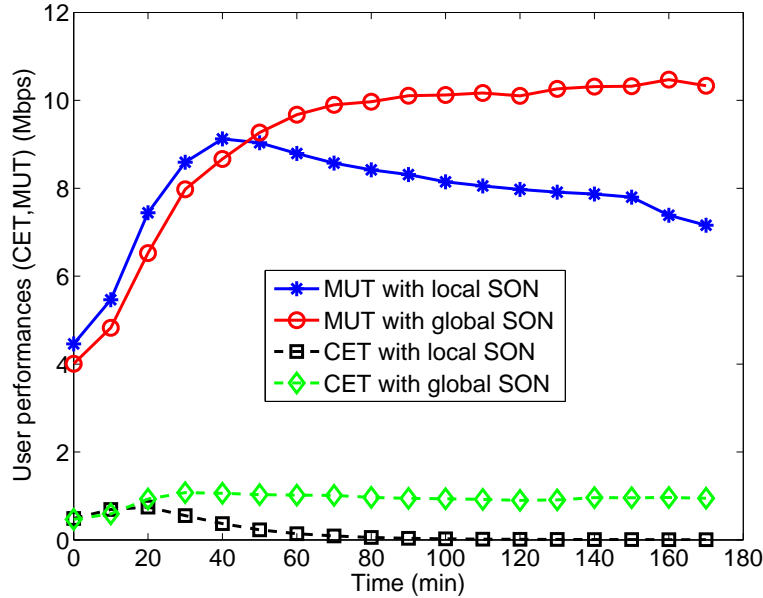}
\caption{ Time evolution of \ac{MUT} and \ac{CET} for the cluster of
the macro BS and the 4 small cells} \label{fig:perfs}
\end{figure}

The overall user performance in terms of \ac{MUT} and \ac{CET}
(see Fig. \ref{fig:perfs}) also
shows the superior performance of Global SON. At the beginning
of the simulation period, the \ac{MUT} is driven by the macro users
which are more numerous. With the activation of the \ac{LB}
algorithms, the macro cell is progressively offloaded by the small cells,
thus the \ac{MUT} improves for both the local SON and the Global SON.
When the real loads are balanced, the global SON stops increasing the
small cells coverage thus ensuring that the \ac{MUT} remains good. The local SON
on the other hand continues to increase the small cells coverage in
order to balance the scheduler loads. This leads to the backhaul 
saturation of certain small cells which see their performance degrade 
drastically and consequently, to the degradation of the overall \ac{MUT}.
The same behavior is observed for the \ac{CET} but this time the performance
degradation for the local SON occurs earlier because cell edge users are more
impacted by an overload in the system. It is noted that the proposed algorithm
\eqref{eq:lb_global} has been proven robust to non-stationary traffic
demands through extensive simulations.

\section{Conclusion} \label{sec:conclusion}
This paper has presented the impact of finite backhaul capacity on
the performance of \ac{LB} in the \ac{RAN}. It has been shown
that by properly defining load indicators that take into account
both the \ac{BS} and the backhaul loads, one can guarantee the
proper operation of the \ac{LB} algorithm. If one neglects
the finite backhaul capacity, the offloading \ac{BS} can saturate. A
load estimator based on measurements has been proposed that takes
 into account the backhaul state. Simulation results of load
balancing in a heterogeneous network with small cells and limited
backhaul capacity have illustrated the importance of using the
correct load definition to avoid possible performance
deterioration. The impact of finite backhaul capacity on other
\ac{SON} algorithms such as \ac{MRO} or \ac{ICIC} would be a natural
extension of this work.

\bibliographystyle{IEEEtran}
\bibliography{main}

\begin{thebibliography}{10}
\providecommand{\url}[1]{#1}
\csname url@samestyle\endcsname
\providecommand{\newblock}{\relax}
\providecommand{\bibinfo}[2]{#2}
\providecommand{\BIBentrySTDinterwordspacing}{\spaceskip=0pt\relax}
\providecommand{\BIBentryALTinterwordstretchfactor}{4}
\providecommand{\BIBentryALTinterwordspacing}{\spaceskip=\fontdimen2\font plus
\BIBentryALTinterwordstretchfactor\fontdimen3\font minus
  \fontdimen4\font\relax}
\providecommand{\BIBforeignlanguage}[2]{{%
\expandafter\ifx\csname l@#1\endcsname\relax
\typeout{** WARNING: IEEEtran.bst: No hyphenation pattern has been}%
\typeout{** loaded for the language `#1'. Using the pattern for}%
\typeout{** the default language instead.}%
\else
\language=\csname l@#1\endcsname
\fi
#2}}
\providecommand{\BIBdecl}{\relax}
\BIBdecl

\bibitem{3gpp_evolveduniversalterrestrial_2008a}
{3GPP}, ``{Evolved Universal Terrestrial Radio Access Network (E-UTRAN);
  Self-configuring and self-optimizing network (SON) use cases and
  solutions},'' {3GPP}, TR 36.902, Sep. 2008.

\bibitem{khan_centralizedLB_2013}
Y.~Khan, B.~Sayrac, and E.~Moulines, ``{Centralized self-optimization of pilot
  powers for load balancing in LTE},'' in \emph{Proc. of IEEE PIMRC}, Sep.
  2013, pp. 3039--3043.

\bibitem{combes_selforganizationwireless_2012}
R.~Combes, Z.~Altman, and E.~Altman, ``{Self-organization in wireless networks:
  a flow-level perspective},'' in \emph{{Proc. of IEEE INFOCOM}}, 2012.

\bibitem{nasri_handoveradaptation_2013}
R.~Nasri and Z.~Altman, ``{Handover adaptation for dynamic load balancing in
  3GPP Long Term Evolution systems},'' in \emph{Proc. of MoMM}, Dec. 2007.

\bibitem{rodriguez_loadbalancing_2011}
J.~Rodriguez, I.~de~la Bandera, P.~Munoz, and R.~Barco, ``{Load balancing in a
  realistic urban scenario for LTE networks},'' in \emph{Proc. of IEEE VTC
  Spring}, 2011, pp. 1--5.

\bibitem{small_cell_forum_BH}
{Small Cell Forum 5.1}, ``Backhaul technologies for small cells,'' \emph{White
  Paper}, Feb. 2014.

\bibitem{osseiran2014scenarios}
A.~Osseiran \emph{et~al.}, ``{Scenarios for 5G mobile and wireless
  communications: the vision of the METIS project},'' \emph{IEEE Communications
  Magazine}, vol.~52, no.~5, pp. 26--35, 2014.

\bibitem{bejerano_fairness_wlan_2007}
Y.~Bejerano, S.-J. Han, and L.~Li, ``Fairness and load balancing in wireless
  lans using association control,'' \emph{IEEE/ACM Transactions on Networking},
  vol.~15, no.~3, pp. 560--573, June 2007.

\bibitem{lee_comp_2012}
D.~Lee \emph{et~al.}, ``{Coordinated multipoint transmission and reception in
  LTE-advanced: deployment scenarios and operational challenges},'' \emph{IEEE
  Communications Magazine}, vol.~50, no.~2, pp. 148--155, Feb. 2012.

\bibitem{ghimire_impactbackhaul_2014}
J.~Ghimire and C.~Rosenberg, ``{Impact of limited backhaul capacity on user
  scheduling in heterogeneous networks},'' in \emph{Proc. of IEEE WCNC}, Apr.
  2014, pp. 2480--2485.

\bibitem{bonald_wirelessdownlinkdata_2003a}
T.~Bonald and A.~Prouti\`{e}re, ``{Wireless Downlink Data Channels: User
  Performance and Cell Dimensioning},'' in \emph{{Proc. of ACM Mobicom}}, 2003.

\bibitem{tall_selforganizingstrategies_2014}
A.~Tall, Z.~Altman, and E.~Altman, ``{Self organizing strategies for enhanced
  ICIC (eICIC)},'' in \emph{{Proc. of WiOpt}}, May 2014, pp. 318--325.

\bibitem{3gpp_evolveduniversalterrestrial_2010}
{3GPP}, ``{Evolved Universal Terrestrial Radio Access (E-UTRA); Further
  advancements for E-UTRA physical layer aspects},'' {3GPP}, TS 36.814 v9.0.0,
  Mar. 2010.

\end{thebibliography}

\end{document}